# More Effective Centrality-Based Attacks on Weighted Networks

Balume Mburano [1], Weisheng Si [1], Qing Cao [2], and Wei Xing Zheng [1]

[1] School of Computer, Data and Mathematical Sciences, Western Sydney University, Sydney, Australia

[2] Department of EECS, University of Tennessee, Knoxville, USA

[1] {b.mburano, w.si, w.zheng}@westernsydney.edu.au, [2] qcao1@utk.edu

*Abstract*—Only when understanding hackers' tactics, can we thwart their attacks. With this spirit, this paper studies how hackers can effectively launch the so-called 'targeted node attacks', in which iterative attacks are staged on a network, and in each iteration the most important node is removed. In the existing attacks for weighted networks, the node importance is typically measured by the centralities related to shortest-path lengths, and the attack effectiveness is also measured mostly by length-related metrics. However, this paper argues that flows can better reflect network functioning than shortest-path lengths for those networks with carrying traffic as the main functionality. Thus, this paper proposes metrics based on flows for measuring the node importance and the attack effectiveness, respectively. Our node importance metrics include three flow-based centralities (flow betweenness, current-flow betweenness and current-flow closeness), which have not been proposed for use in the attacks on weighted networks yet. Our attack effectiveness metric is a new one proposed by us based on average network flow. Extensive experiments on both artificial and real-world networks show that the attack methods with our three suggested centralities are more effective than the existing attack methods when evaluated under our proposed attack effectiveness metric.

## I. INTRODUCTION

Infrastructure networks such as the Internet Backbone, power grids and wireless mesh networks are critical for our society and economy. Due to their high importance and wide presence, they have been the hot targets of cyber-attacks [1]. Infamous incidents of such attacks include the 'WannaCry' attack on the UK's National Health Service network in 2017 [2], the ransomware attack on USA Colonial Pipeline's oil pipelines in 2021 [3], etc.

With the prevalence of cyber-attacks, the best strategy to resist them is to study how hackers think. Following this strategy, this paper investigates how hackers can effectively launch a type of attacks called the 'targeted node attacks', in which rounds of attacks are carried out on a network, and in each round the most important node and its incident links are removed from the network, until no nodes are left [4-6]. Here the node importance can be measured by any metric the hackers choose, e.g., node degree, various kinds of centralities, node clustering coefficient, etc.

In calculating those node importance metrics, there are two approaches [7]: *initial* and *adaptive*. In the *initial* approach, the metric values of all nodes are calculated based on the initial network only, and in each round of attack, the initial values are used to select the node to remove. In the *adaptive* approach, the metric values of remaining nodes are recalculated after each round of attack. It was shown in [7] that the attacks with the adaptive approach are more effective than those attacks with the initial approach. Therefore, most research works propose the attack methods with the adaptive approach, and so will this paper.

Besides being initial or adaptive, another aspect of attacks is whether to consider the link weights in the networks. This aspect affects the measurements of both node importance and attack effectiveness.

On one hand, most of the existing attack methods (e.g., [5, 6, 8-11]) consider the links unweighted, thus achieving low computation complexity. In these methods, the node importance is measured by those unweighted metrics such as node degree, unweighted betweenness centrality [12], unweighted closeness centrality [12], etc. The attack effectiveness is measured by a metric called $R$ [4] or its variants. $R$ was inspired by the Percolation Theory [13], in which the size of the Largest Connected Component (LCC) in a network is the key indicator on network connectivity. We give the formula for $R$ in Section II. Note that $R$ is originally proposed as a metric for measuring network robustness, not for attack effectiveness. However, since higher network robustness means less attack effectiveness, $R$ is used to measure attack effectiveness as well, with a less value of $R$ indicating a more effective attack.

On the other hand, since the links in most infrastructure networks have some kinds of weights (e.g., capacity, length, etc.), an attack method considering link weights can be more destructive. Consequently, several weighted attack methods [14-17] have appeared in recent years. In these methods, the node importance is measured by node strength (the total weight of the links incident to a node) (e.g., in [14]), weighted betweenness centrality (in [15, 17]), weighted closeness centrality (in [15, 17]), etc. Note that weighted betweenness and closeness are based on shortest-path lengths and hence called shortest-path betweenness and closeness as well [12]. They will be detailed in Section II. In terms of attack effectiveness, these methods measure it by $R$ (e.g., in [14]), Network Efficiency (EFF) (in [14, 15]), Total Flow (in [15]), Average Shortest Path Length

(ASPL) (in [17]), etc. Note that EFF and ASPL are based on shortest-path lengths and will be detailed in Section II.

It can be seen from the above that most existing weighted attack methods reference shortest-path lengths for measuring node importance and attack effectiveness. However, shortest-path lengths suffer from two weaknesses: (1) not all paths are considered in a network, and (2) the capacity for carrying network traffic is not considered. To overcome these two weaknesses, this paper proposes new attack methods with metrics based on *flows* (including both network flows [18] and electrical currents [19]) for measuring node importance and attack effectiveness. Specifically, this paper makes the following contributions:

- Flow betweenness centrality [20], current-flow betweenness centrality [21] and current-flow closeness centrality [19] are proposed as the node importance metrics for use in the weighted attacks. All of them consider flows instead of shortest-path lengths in calculating the centrality values.

- A new metric for measuring attack effectiveness called $R_{ANF}$ is proposed. $R_{ANF}$ is a variation to $R$ by replacing the LCC size with the Average Network Flow (ANF) proposed in our previous work [22]. Since ANF considers the flows among all node pairs in a network, $R_{ANF}$ is a fine-grained metric for measuring attack effectiveness.

- Extensive experiments on both artificial and real-world networks are conducted, showing that the three centralities suggested above lead to more effective attack methods than the existing ones when using $R_{ANF}$ as the attack effectiveness metric. Among these three centralities, the current-flow betweenness results in the most effective attack method.

The rest of this paper is arranged as follows. Section II introduces the preliminaries for this paper. Section III discusses related works and points out their weaknesses. Section IV details our attack methods with the metrics for measuring node importance and for measuring attack effectiveness. Section V and Section VI describe the experimental setup and results. Finally, Section VII concludes this paper.

## II. PRELIMINARIES

This section introduces the graph notations used in this paper and discusses the metrics for node importance and attack effectiveness.

This paper models a network by an undirected graph $G = (V, E)$, where $V$ is the set of vertices (or nodes) and $E$ is the set of links (or edges). For convenience, we let $n = |V|$ and $m = |E|$, and label the nodes in $V$ from 1 to $n$. An edge connecting nodes $i$ and $j$ is denoted by $(i, j)$. The weight associated with edge $(i, j)$ is denoted by $c(i, j)$. The assumption of undirected graphs in this paper is because most of the infrastructure networks support bidirectional communications on their links.

### A. Metrics for Node Importance

The first three metrics in this subsection are the main ones used in the existing works, and the last three are proposed for use in the attacks by this paper.

#### 1) Node Strength (NS):

The NS of a node $i$, denoted by NS($i$), is the total weight of all edges incident to node $i$ [14]. The formula for calculating NS($i$) is as follows:

$$\text{NS}(i) = \sum_{v \in N(i)} c(i, v) \quad (1)$$

Here $N(i)$ denotes the set of neighbor nodes to node $i$.

#### 2) Weighted Betweenness Centrality:

The weighted (shortest-path) betweenness centrality for a node $i$, denoted by $C_{SPB}(i)$, reflects the chance that node $i$ is on the shortest paths among node pairs in a network [17]. The formula for calculating $C_{SPB}(i)$ is as follows:

$$C_{SPB}(i) = \frac{2}{(n-1)(n-2)} \sum_{s,t \in V-\{i\}, s<t} \frac{\tau(s,t \mid i)}{\tau(s,t)} \quad (2)$$

Here $s$ and $t$ are two distinct nodes in the network other than node $i$. The '$s < t$' is used to indicate that a node pair $(s, t)$ is only counted once. The $\tau(s,t)$ represents the total number of shortest paths between $s$ and $t$, and the $\tau(s,t \mid i)$ represents the number of shortest paths between $s$ and $t$ that pass through $i$. Since we have totally $(n-1)(n-2)/2$ node pairs in $V - \{i\}$, dividing it gives the average over all node pairs.

Note that, for the unweighted betweenness centrality, the formula remains unchanged, but the lengths of all links are simply deemed as '1'. The same way is applied to the weighted closeness centrality below to get its unweighted version.

#### 3) Weighted Closeness Centrality

The weighted (shortest-path) closeness centrality for a node $i$, denoted by $C_{SPC}(i)$, is the reciprocal of the average distance from node $i$ to all other $n-1$ nodes in a network [17]. The formula for calculating $C_{SPC}(i)$ is as follows:

$$C_{SPC}(i) = \frac{n-1}{\sum_{i \in V} d(i,v)} \quad (i \neq v) \quad (3)$$

Here $d(i,v)$ gives the shortest path distance between node $i$ and node $v$. Basically, the weighted closeness centrality reflects the average distance between a node and all the other nodes in a network. Thus, the closer the node $i$ is to all other nodes, the more central the node $i$ is.

#### 4) Flow Betweenness Centrality

The flow betweenness [20] is based on the maximum flows among node pairs in a flow network [18]. The flow betweenness for a node $i$, denoted by $C_{FB}(i)$, is calculated as follows:

$$C_{FB}(i) = \frac{2}{(n-1)(n-2)} \sum_{s,t \in V-\{i\}, s<t} \frac{F(s,t \mid i)}{F(s,t)} \quad (4)$$

Here $s$ and $t$ are two distinct nodes in the network other than node $i$; $F(s, t)$ represents the maximum flow between $s$ and $t$; and $F(s, t \mid i)$ represents the maximum flow between $s$ and $t$ that pass through $i$; and dividing $(n - 1)(n - 2)/2$ gives the average over all node pairs in $V - \{i\}$.

*5) Current-Flow Betweenness Centrality*

The current-flow betweenness [21] is based on the electrical currents among node pairs in an electrical circuit. The difference between a network flow and an electrical current for a node pair is that a network flow (even when achieving the maximum) may abandon some paths between this node pair, while an electrical current always uses all the possible paths [21]. Specifically, the Current-Flow Betweenness Centrality for node $i$, denoted by $C_{CFB}(i)$, is calculated as follows:

$$C_{CFB}(i) = \frac{2}{(n-1)(n-2)} \sum_{s,t \in V-\{i\}, s<t} I_{s,t}(i) \quad (5)$$

Here $I_{s,t}(i)$ gives the electrical current passing through node $i$ when a unit current is applied between $s$ and $t$.

*6) Current-Flow Closeness Centrality*

The current-flow closeness [19] also views a network as an electrical circuit and uses the potential difference (i.e., voltage) between two nodes as the closeness measure. Specifically, the Current-Flow Closeness Centrality for a node $i$, denoted by $C_{CFC}(i)$, is calculated as follows:

$$C_{CFC}(i) = \frac{n-1}{\sum_{i \in V} PD(i, v)} \quad (i \neq v) \quad (6)$$

Here $PD(i, v)$ is the potential difference between node $i$ and node $v$ when a unit current is applied between them.

Note that, unlike 'flow betweenness centrality', 'flow closeness centrality' does not exist in the literature, since a flow network has no concepts similar to the potential difference.

*B. Metrics for Attack Effectiveness*

This subsection covers four main metrics used for measuring attack effectiveness in the existing weighted attack methods.

*1) R*

$R$ is the average of the ratios of the LCC size and the network size after all rounds of attacks [4]. Here 'size' means the number of nodes. Specifically, given a network $G$, the formula for calculating $R$ is as follows:

$$R(G) = \frac{1}{n} \sum_{i=1}^{n} \frac{LCC(G_i)}{n} \quad (7)$$

Here $G_i$ denotes the network after the $i$-th round of attack, and $LCC(G_i)$ denotes the LCC size in $G_i$, and the $1/n$ at the beginning averages the results of all rounds.

Note that although $R$ has been used in some weighted attacks (e.g., [14, 15]) to measure attack effectiveness, this is not suitable since the LCC size does not reflect link weight. However, $R$ provides a template formula for the remaining three metrics discussed in this subsection. These three metrics are used to replace the LCC size in the formula (7) to obtain the final metric. That is, the average result of a metric of all attack rounds serves as the final metric.

*2) Total Flow (TF)*

TF is the sum of the capacities of all links in a graph $G$ [15]. Given a graph $G$, the formula for calculating TF is as follows:

$$TF(G) = \sum_{(i,j) \in E} c(i, j) \quad (8)$$

Note that, albeit the name of TF has 'flow' in it, it does not consider the flows [18] among node pairs, but simply the link capacities. Thus, it is much less fine-grained than ANF, which considers the flows among all node pairs in a network.

*3) Average Shortest Path Length (ASPL)*

ASPL is the average of the shortest path lengths of all node pairs in a graph $G$ [17]. It can be used to measure attack effectiveness because a network with a smaller ASPL is deemed to have better connectivity. Given a network $G$, the formula for calculating ASPL is as follows:

$$ASPL(G) = \frac{2}{n(n-1)} \sum_{s,t \in V, \ s<t} d(s, t) \quad (9)$$

Here $d(s, t)$ denotes the shortest path length between node $s$ and node $t$. Since there are totally $n(n-1)/2$ node pairs in a graph, dividing it gives the average over all node pairs.

*4) Network Efficiency (EFF)*

EFF is the average of the reciprocals of shortest path lengths of all node pairs in a graph $G$ [15]. It can be used to measure attack effectiveness because a network with a smaller EFF is deemed to have less delay (i.e., better efficiency). Given a network $G$, the formula for calculating EFF is as follows:

$$EFF(G) = \frac{2}{n(n-1)} \sum_{s,t \in V, \ s<t} \frac{1}{d(s, t)} \quad (10)$$

### III. RELATED WORK

This section surveys the existing weighted attack methods and points out their weaknesses by discussing the following two aspects: how to measure node importance and how to measure attack effectiveness.

*A. How to Measure Node Importance*

In [14, 15], Node Strength (NS) and weighted betweenness centrality were proposed as the node importance metrics for the targeted node attacks. Since NS is a local metric only considering the incident links of a node, the attack method with NS is shown to be less effective than the attack method with weighted betweenness, which is a global metric considering all other nodes in a network. In [17], both weighted betweenness and closeness were proposed to measure node importance. It was found that the attack methods with both of them are more effective than the attack method with NS, and the attack method with weighted betweenness is slightly better than the attack method with weighted closeness. However, since both weighted betweenness and weighted closeness are based on shortest-path length, we show later in this paper that our suggested metrics based on flows give more effective attack methods.

Moreover, the recent [16] proposed to use the *conditional weighted betweenness centrality* to measure node importance. Its basic idea is to only calculate the weighted betweenness for those nodes in the LCC and then select the node with the highest betweenness value. A similar idea was independently proposed in [9], but it was called the 'the largest component strategy' there and was applied to all node importance metrics for unweighted networks. Later, it was proved in [6] that the largest component strategy is a necessary condition for minimizing the $R$ value (i.e., the best attack) for unweighted networks. However, since the largest component strategy resorts to the LCC that does not depend on link weight, its application to weighted networks does not always give the most effective attack [16]. Thus, our experiments did not compare with [16].

*B. How to Measure Attack Effectiveness*

In [14], the LCC size and the EFF are integrated into the $R$'s template formula (7) to get the attack effectiveness metrics. However, the LCC size is not related to link weight, so it is not very suitable in the weighted attack scenario. As discussed in the previous section, the EFF is based on shortest-path length and reflects the delays in a network, so it is not closely related to the traffic-carrying capability of the network. In addressing the shortcoming of EFF, the TF metric (which is based on link capacity) was proposed [15]. However, TF only calculates the sum of the capacities of all the links and does not consider the flows among the node pairs in the network, so it is less fine-grained than the ANF proposed by us. In the quite recent [17], the ASPL (also called the 'average geodesic distance' in that paper) was proposed for measuring attack effectiveness. Unfortunately, it suffers from the same weakness as EFF since it is based on shortest-path length.

## IV. OUR PROPOSED ATTACK METHODS

This section first discusses the three node importance metrics proposed for use in weighted attacks, and then details the steps of the attack methods, and finally presents our new metric $R_{ANF}$ for measuring attack effectiveness.

*A. Our Metrics for Measuring Node Importance*

We propose to use the following three centralities to select the node for removal in the targeted node attacks.

- *Flow Betweenness Centrality* ($C_{FB}$): As suggested by its definition in Section II, the flow betweenness is better than the shortest-path betweenness for weighted networks because the flows consider not only multiple paths between a node pair but also the link capacities on these paths. The number of paths contributing to the maximum flows is much more than that of the shortest paths.
- *Current-Flow Betweenness Centrality* ($C_{CFB}$): As discussed in Section II, the current-flow betweenness improves over the flow betweenness by considering all the possible paths between a node pair, while the flow betweenness may oddly ignore some paths between a

**Algorithm 1:** Targeted Attacks with Proposed Centralities

1. Calculate the connected components of the current network.
2. Calculate the centrality value for each node in the connected component where it belongs.
3. Remove the node with the highest centrality value among the entire network. Ties are broken randomly.
4. Go to Step 1 if there are still nodes in the network.

    node pair even when achieving the maximum flow [21]. Thus, we propose current-flow betweenness for use in the attacks as well.
- *Current-Flow Closeness Centrality* ($C_{CFC}$): Besides current-flow betweenness, current-flow closeness is another metric considering electrical currents in a network. Thus, it is interesting to see how it performs as a node importance metric in the attacks.

As noted in Section II, since the concept of flow closeness centrality is not feasible and does not exist, it is absent in this paper.

*B. The Steps of Attack Methods*

Most infrastructure networks have the concept of link capacity. For example, the Internet backbone has bandwidth, and the power grid has the voltage (usually very high voltage above 50 KV to reduce power loss), etc. In calculating $C_{FB}$, we directly use the link capacities stated in the dataset for a network. In calculating $C_{CFB}$ and $C_{CFC}$, a critical question is how to translate the capacity of a link to the resistance of this link, while maintaining the definitions of $C_{CFB}$ and $C_{CFC}$ where a unit electrical current is applied between a node pair. Here our solution is: *given a link capacity (whatever it is — bandwidth, voltage, etc.), we deem it as the conductance of this link, and then calculate the resistance of this link, which equals the reciprocal of conductance*. With this interpretation, we can use the existing algorithm proposed in [19] to calculate both $C_{CFB}$ and $C_{CFC}$.

As a note, in our previous work [9], $C_{CFB}$ and $C_{CFC}$ have been proposed to measure node importance for unweighted networks. However, their use there does not involve the above solution, since the resistance of each link is simply deemed as '1' for unweighted networks.

Now we are ready to describe the steps of the targeted attacks with the proposed centralities in the Algorithm 1 and provide the following explanations.

- If a node importance metric (e.g., Node Strength) can be calculated on a disconnected network, then Step 1 is not needed. However, for our proposed $C_{FB}$, $C_{CFB}$ and $C_{CFC}$, and the existing $C_{SPB}$ and $C_{SPC}$ to be compared with, their values for a node need to be calculated within the connected component where the node resides.

- In Step 2, the centrality measure can be any of $C_{FB}$, $C_{CFB}$, $C_{CFC}$, $C_{SPB}$, and $C_{SPC}$. There are existing polynomial-time algorithms for calculating them [12, 19].

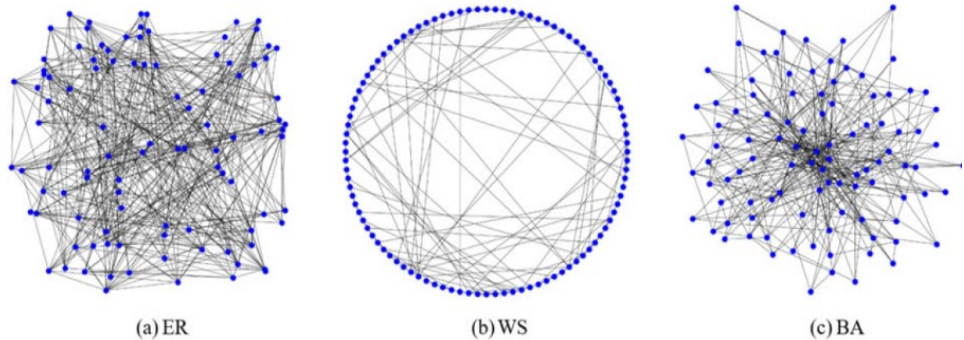

Fig. 1. Exemplar networks from the three well-known models for Complex Networks

*C. Our Metric for Measuring Attack Effectiveness*

The basic idea of our metric $R_{ANF}$ is to leverage the ANF proposed in our previous work [22] to measure the flow-carrying capability of the networks after attacks, thus reflecting the attack effectiveness. Below we first recall the definition of ANF, and then present the formula for calculating $R_{ANF}$.

Given a network $G=(V, E)$ with each edge having a non-negative capacity, the ANF on $G$, denoted by ANF($G$), is the average of the maximum flows among all node pairs in $G$. The calculation formula is as follows:

$$ANF(G) = \frac{2}{n(n-1)} \sum_{s,t \in V,\ s<t} F(s,t) \quad (11)$$

Here $F(s, t)$ denotes the maximum flow between two nodes $s$ and $t$. Since there are altogether $\frac{n(n-1)}{2}$ node pairs in $G$, the averaging factor is $\frac{2}{n(n-1)}$. Note that efficient algorithms [22] exist such that there is no need to calculate each individual $F(s, t)$ to get their average.

Our metric $R_{ANF}$ calculates the average of ANFs over all attack rounds. Its formula is given as follows:

$$R_{ANF}(G) = \frac{1}{n} \sum_{i=1}^{n} \frac{ANF(G_i)}{ANF(G)} \quad (12)$$

Here $G_i$ denotes the network after the $i$-th round of attack and $G$ denotes the initial network. The ratio ANF($G_i$)/ANF($G$) is calculated in each round. The $1/n$ is the averaging factor.

Since the ratio of ANF($G_i$)/ANF($G$) is always less than 1, it enables $R_{ANF}$ to have the following two nice properties: (1) *it will not be unfairly affected by the network size and the scale of link capacity*; (2) *its value ranges between 0 and 1*.

Based on formula (12), $R_{ANF}$ can be computed by obtaining ANF($G_i$)/ANF($G$) after the Step 3 in Algorithm 1, and then averaging this ratio over all attack rounds.

## V. EXPERIMENTAL SETUP

Our experiments use both artificial networks and real-world networks as attack targets. This section describes the generation of artificial networks and the datasets of real-world networks.

*A. Artificial Networks*

The following three well-known models for Complex Networks [12] are used to generate artificial networks.

- The Erdos-Renyi (ER) model [23] for random networks.
- The Barabasi-Albert (BA) model [24] for scale-free networks.
- The Watts-Strogatz (WS) model [25] for small-world networks.

Fig. 1 depicts the exemplar networks from these three models. We used Python's NetworkX package [26], which include the network generators for these three models, for the network generation. In addition, the link capacities are generated as random integers between 1 and 10.

*B. Real-world Networks*

The real-world networks used in our experiments include the following two power grids. The datasets of both of them contain the data about network topologies and link capacities.

- *Australian Electricity Power Grid* [27]: it comprises 1529 nodes and 2377 links. The nodes represent the power stations, and the links represent the power lines between the stations. The link capacities are given in the form of voltage which ranges from 66KV to 500KV. Note that the voltages in the power lines are typically very high to reduce energy loss.

- *European Electrical Transmission Network* [28]: it comprises 1479 nodes and 2289 links. The link capacities are given by two fields: voltage (ranging from 220KV to 500KV) and the number of cables (ranging from 1 to 9). Here we use the voltage times the number of cables as link capacity.

## VI. EXPERIMENTAL RESULTS

Using the metric $R_{ANF}$, we measured the attack effectiveness of our proposed methods (with $C_{FB}$, $C_{CFB}$ and $C_{CFC}$ as node importance metrics) and the existing methods (with NS, $C_{SPB}$ and $C_{SPC}$ as node importance metrics). These attack methods are implemented by Python's NetworkX package [26].

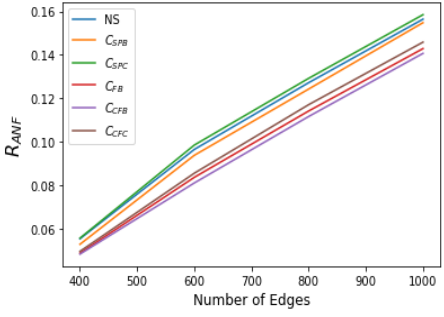
Fig. 2. Attack effectiveness on BA networks with different $m$'s and $n=200$

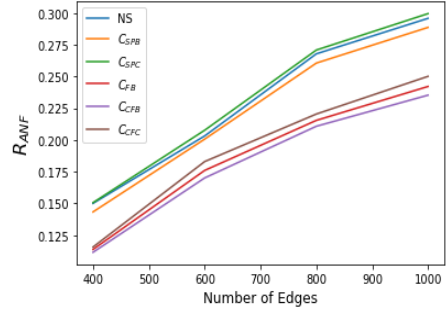
Fig. 3. Attack effectiveness on ER networks with different $m$'s and $n=200$

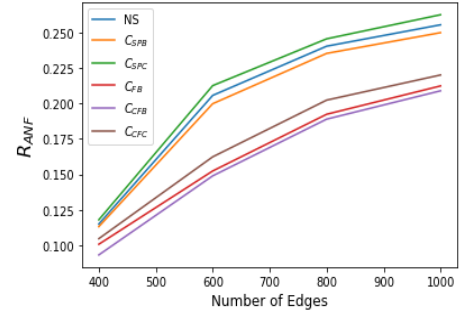
Fig. 4. Attack effectiveness on WS networks with different $m$'s and $n=200$

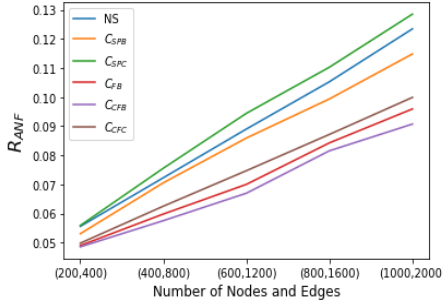
Fig. 5. Attack effectiveness on BA networks with different $(n,m)$'s

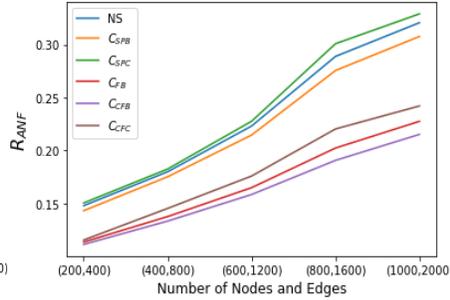
Fig. 6. Attack effectiveness on ER networks with different $(n,m)$'s

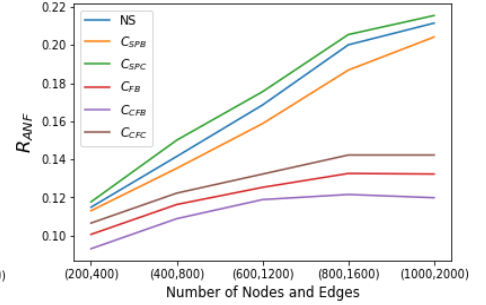
Fig. 7. Attack effectiveness on WS networks with different $(n,m)$'s

### A. Attack Effectiveness on Artificial Networks

Two groups of experiments are conducted on all three models of artificial networks.

- In the first group, we fix $n=200$ and vary $m$ to be 400, 500, …, 1000 respectively. This is for examining the impact of edge density on the $R_{ANF}$ values.

- In the second group, we vary both $n$ and $m$, but maintain the $m/n$ ratio as 2: $n=200, 400, …, 1000$, and $m=400, 800, …, 2000$ respectively. This is for examining the impact of network scale.

For each $(n, m)$ setting under a network model, a set of 50 networks is generated randomly. In the coming figures, a data point represents an average $R_{ANF}$ value from a set of 50 networks.

Figs.2-4 present the results from the first group of experiments. From these figures we can observe that:

- The three metrics suggested by us give more effective attacks than the existing ones. This is demonstrated by the $R_{ANF}$ values of $C_{FB}$, $C_{CFB}$ and $C_{CFC}$ being smaller than those of NS, $C_{SPB}$ and $C_{SPC}$. In particular, our suggested $C_{CFB}$ sees the smallest $R_{ANF}$ values, leading to the most effective attack; and the existing $C_{SPC}$ sees the largest $R_{ANF}$ values, leading to the least effective attack.

- For all six attack methods, the $R_{ANF}$ values increase with edge density. Thus, even though the definition of $R_{ANF}$ includes normalization to confine the impact of network scale by making $R_{ANF} \in [0,1]$, the network scale still shows some impact on $R_{ANF}$. This is because when the network scale is larger, the early attack rounds will see the ANF$(G_i)$/ANF$(G)$ ratios closer to 1, which enlarges the final value of $R_{ANF}$.

Figs.5-7 present the results from the second group of experiments. These results are consistent with those in the first group. Basically, they also show that:

- The three metrics suggested by us result in more effective attack methods than the existing ones. In particular, our suggested $C_{CFB}$ leads to the most effective attack, and the existing $C_{SPC}$ leads to the least effective attack.

- For all six attack methods, the $R_{ANF}$ values increase with network scale.

Overall, these two groups of experiments confirm our idea that flow-based centralities can lead to more effective attacks than those metrics based on shortest paths. Moreover, with $C_{CFB}$ considering all the possible paths between a node pair, it gives us the best attack.

### B. Attack Effectiveness on Real-World Networks

This subsection presents the experimental results on real-world networks. Fig. 8 and Fig. 9 plot the $R_{ANF}$ values of the six attack methods on the Australian Power Grid and the European Electrical Network, respectively. The results from both figures are consistent with Figs. 2-7. That is, the three metrics suggested by us give rise to more effective attacks than the existing ones. In particular, our suggested $C_{CFB}$ gives the most effective attack, and the existing $C_{SPC}$ gives the least effective attack.

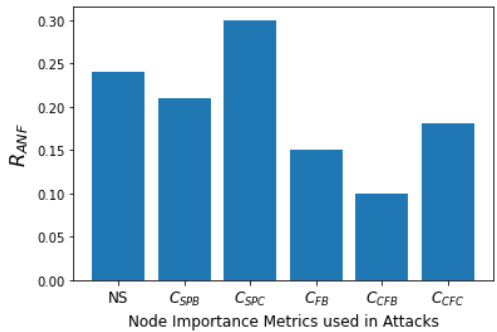

Fig. 8. Attack effectiveness on Australian Power Grid

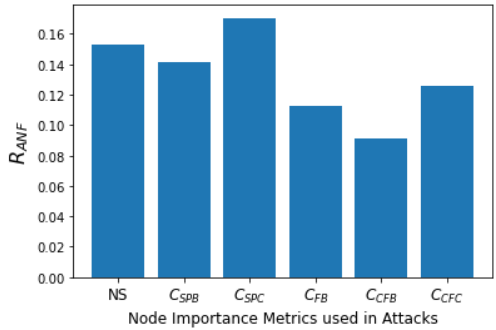

Fig. 9. Attack effectiveness on European Electrical Network

The $R_{ANF}$ values in the Australian Power Grid are generally higher than those in the European Electrical Network. This is because the link capacities in the Australian Power Grid vary less than the link capacities in the European Electrical Network, where the number of cables differentiate the link capacities significantly.

VII. CONCLUSION

This paper is motivated by the observation that the existing weighted attacks mostly use the metrics related to shortest-path lengths to remove nodes. These metrics consider neither the link capacities nor the paths other than the shortest paths when measuring node importance. To overcome these two limitations, this paper proposed to use three flow-based centralities (flow betweenness, current-flow betweenness, and current-flow closeness) as node importance metrics during the attacks. Moreover, this paper proposed a new metric called $R_{ANF}$, which is based on average network flow, for measuring attack effectiveness. Extensive experiments on both artificial and real-world networks showed that, evaluated under $R_{ANF}$, the attack methods with the three centralities suggested by us outperform the existing attack methods, and the attack method with the current-flow betweenness performs the best.